\documentclass[fleqn,10pt]{wlscirep}
\usepackage[utf8]{inputenc}
\usepackage[T1]{fontenc}
\usepackage{amsmath,bm}

\title{Hierarchical team structure and multidimensional localization (or siloing) on networks}

\author[1,2,3,*]{Laurent H\'ebert-Dufresne}
\author[3,4]{Guillaume St-Onge}
\author[1]{John Meluso}
\author[1,5]{James Bagrow}
\author[1,3,4]{Antoine Allard}
\affil[1]{Vermont Complex Systems Center, University of Vermont, Burlington VT}
\affil[2]{Department of Computer Science, University of Vermont, Burlington VT}
\affil[3]{D\'epartement de physique, de g\'enie physique et d'optique, Universit\'e Laval, Qu\'ebec (Qu\'ebec), Canada G1V~0A6}
\affil[4]{Centre interdisciplinaire en mod\'elisation math\'ematique, Universit\'e Laval, Qu\'ebec (Qu\'ebec), Canada G1V~0A6}
\affil[5]{Department of Mathematics \& Statistics, University of Vermont, Burlington VT}
\affil[*]{laurent.hebert-dufresne@uvm.edu}

\begin{abstract}
Knowledge silos emerge when structural properties of organizational interaction networks limit the diffusion of information.
These structural barriers are known to take many forms at different scales --- hubs in otherwise sparse organisations, large dense teams, or global core-periphery structure --- but we lack an understanding of how these different structures interact.
Here we bridge the gap between the mathematical literature on localization of spreading dynamics and the more applied literature on knowledge silos in organizational interaction networks.
To do so, we introduce a new model that considers a layered structure of teams to unveil a new form of hierarchical localization (i.e., the localization of information at the top or center of an organization) and study its interplay with known phenomena of mesoscopic localization (i.e., the localization of information in large groups), $k$-core localization (i.e., around denser $k$-cores) and hub localization (i.e., around high degree stars).
We also include a complex contagion mechanism by considering a general infection kernel which can depend on hierarchical level (influence), degree (popularity), infectious neighbors (social reinforcement) or team size (importance). 
This general model allows us to study the multifaceted phenomenon of information siloing in complex organizational interaction networks and opens the door to new optimization problems to promote or hinder the emergence of different localization regimes.
\end{abstract}
\begin{document}

\flushbottom
\maketitle

\thispagestyle{empty}

\section{Introduction}

Organizational interaction networks, from those that follow the structures of private companies and social media to public institutions and academia, have long been known to influence the spread of information and dynamics among their agents \cite{katz1963traditions, clement2018evolution, bento2020organizational}. Open source projects can falter or thrive through the interplay of maintainers and newcomers \cite{ahuja1999network, long2007social, hinds2008social} much like how social media network structures hasten or hinder the development of echo chambers around specific groups \cite{evans2018opinion, hebert2017strategic}. The collaborative structure in science can both create knowledge silos or help diffuse new ideas \cite{long2013bridges}, just as the hierarchical structure of an organization can shape culture \cite{melusoInPressmasculinity} and individual engagement \cite{mitra2017spread, clement2018evolution}. The strength of connections between individuals further shape these effects. Ideas and emotions spread over weak organizational ties \cite{granovetter1973weakties, macy1990learning, smith-crowe2014corruption}. Stronger ties and repeated exposures amplify these effects \cite{watson2006knowledge, ferrali2020crime}, facilitating the spread of socialized behaviors and actions \cite{centola2007complex, melusoInPressmasculinity}. At a theoretical level, all of these phenomena are related to the idea of dynamic localization: dictating whether specific parts of a system contribute alone to a dynamical process (\textit{localized}), or whether the system as a whole undergoes collective activity (\textit{delocalized}).

``Knowledge silos'' (or \textit{thought worlds} \cite{dougherty1992thoughtworlds}) that constrain information or activity within specific parts of an organisation are neither nor good nor bad per se. They can be detrimental to organisations if they limit adoption of norms and practices and hinder the spread of innovation \cite{majchrzak2011transcending}. They can also be desired when they protect confidential information or foster social reinforcement within teams \cite{wenger2000communities}. In any case, knowing about the potential for different types of silos can be important. Through a better understanding of the different theoretical mechanisms under which silos emerge would help us control and design systems with information localization in mind. We therefore aim to provide a broad framework for localization (the mathematical phenomenon) in order to inform future work around silos (the empirical manifestation).

In an organisational interaction network of nodes (agents) and edges (information pathways), dynamical localization can be defined as a phase transition or an exponentially increasing level of activity across node categories with different levels of connectivity \cite{martin2014localization,st-onge2021social}. For instance, if we consider only degree heterogeneity, simple models generally assume a node with $k$ neighbors to be exposed $k$ times more to some dynamical process than a node with a single neighbor and therefore to be $k$ times more active (or less, if saturation occurs) \cite{pastor2001epidemic}. However, for sufficiently heterogeneous networks, more detailed descriptions show that activity scales exponentially with the level of connectivity of a node \cite{stonge2018phase}. The subsets of the networks where dynamics localize can then be identified in simulations with ad hoc tools or, as we will do, predicted analytically by demonstrating the exponential relationship.

Several approaches exist to identify specific types of localization, resulting in a rich and varied literature. 
By modeling epidemic spread and using a spectral approach~\cite{dorogovtsev2003spectra}, Goltsev \textit{et al}. showed in 2012 how a dynamical process can localize around the hubs of a network~\cite{goltsev2012localization}. In particular, when studying a simple contagion process which diffuses at a normalized rate $\lambda$ and dissipates at a unit rate, one can define a local activation threshold of a star subgraph (or star motif) with $k$ neighbors as $\lambda_i \sim 1/\sqrt{k_i}$. Of course, in a larger complex network, the local activation threshold would be decreased by other active nodes in the system. Still, it remains that in heterogeneous networks, the node with the largest degree, $k_{\textrm{max}}$, can maintain activity on its own if the transmission rate lies around $1/\sqrt{k_{\textrm{max}}}$, which is often much smaller than the rate necessary for the collective activation of the whole network. This phenomenon can also be captured numerically by studying the participation ratios of individual nodes to either the dynamical steady state or the eigenvector centrality of the whole system \cite{martin2014localization}.

\begin{figure}
    \centering
    \includegraphics[width=0.88\linewidth]{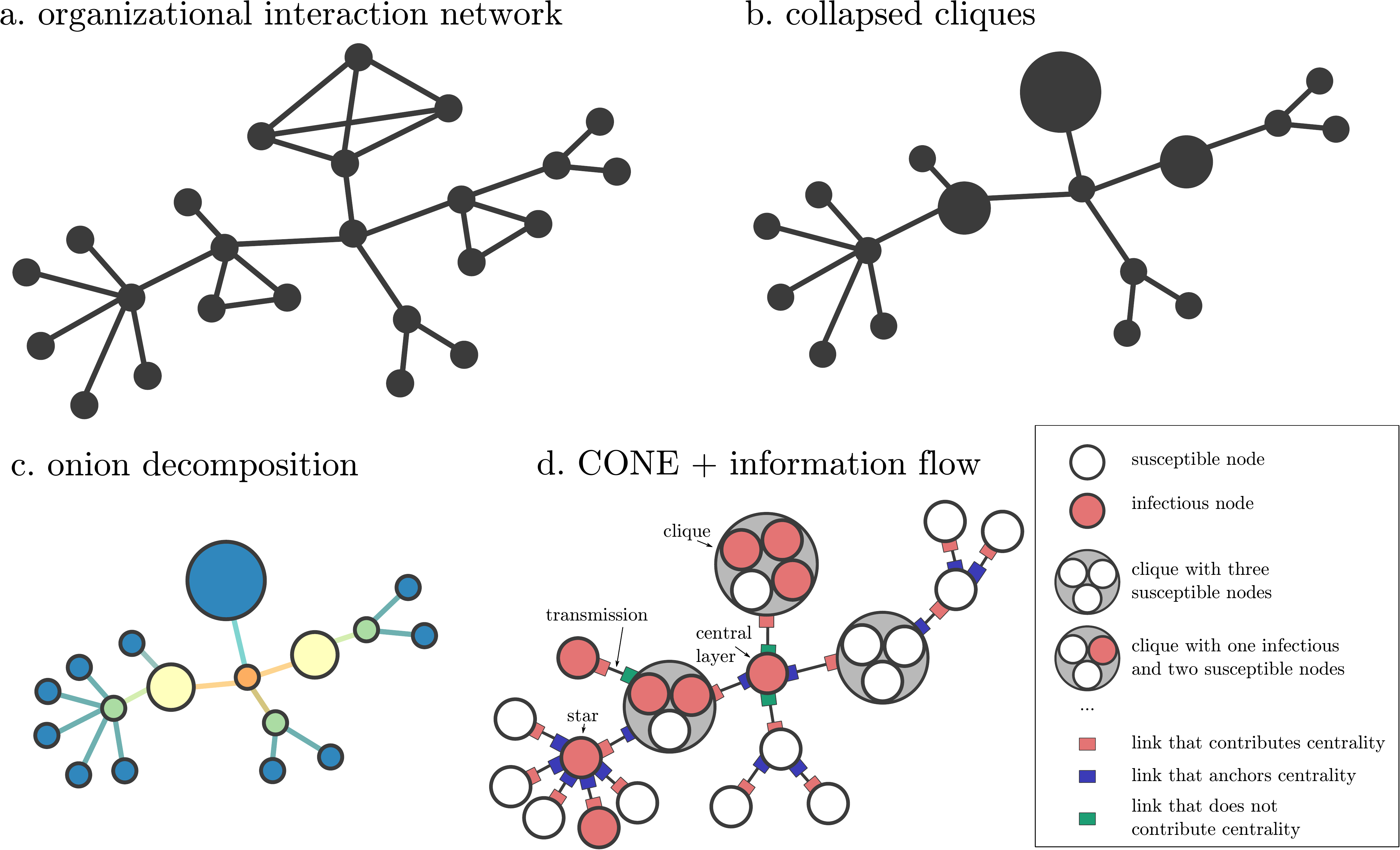}
    \caption{\textbf{Key structural properties for dynamical localization and for the Clustered Onion Network Ensemble.} \textbf{(a.)} A small example of an organizational interaction network featuring a mixture of hubs, cliques and central nodes. \textbf{(b.)} We collapsed cliques on teams (or metanodes). \textbf{(c.)} We classify the hierarchy of the resulting team network using the onion decomposition, where node colors represent centrality layers: layer 1 in blue, 2 in green, 3 in yellow, 4 in orange. \textbf{(d.)} Our model captures the resulting structure and dynamics through the degree $k$, onion layer $\ell$, and size $n$ of nodes and teams. Information silos can be found around stars (nodes of large degree), levels of a hierarchy (central layers) or cliques (large teams). 
    }
    \label{fig:cartoon}
\end{figure}

The description of the localization phenomenon was extended beyond degree (a \textit{microscopic} metric of connectivity) to $k$-core centrality (a \textit{macroscopic} metric of connectivity) in 2018 by Pastor-Satorras and Castellano~\cite{pastor-satorras2018eigenvector}. Based on the $k$-core decomposition --- which prunes the network to identify the nested maximal subsets (or $k$-cores) of nodes who have at least $k$ connections among each other --- this version of localization lets the denser core of a network maintain a spreading process on its own with subcritical spillovers in the periphery of the network.
More recently, St-Onge \textit{et al.} investigated the possibility of \textit{mesoscopic} localization around specific groups rather than nodes or cores~\cite{st-onge2021social, st-onge2021master}. 
Finding that a heterogeneous distribution of group size and weak (finite) coupling could allow large enough groups to maintain a diffusion process, again with subcritical spillover into smaller groups.
These results all show that diverse sets of subgraphs with high enough connectivity can localize and maintain dynamics independently, despite the lack of collective activation across the entire network. We draw an example of the structural network properties relevant to visualization in Fig.~\ref{fig:cartoon}, illustrating how the interplay of degree, centrality, and group density, can be hard to quantify even in simple examples.

In this paper, we pursue a holistic view of dynamical localization in networks. Are there other mechanisms for localization which have been so far ignored? In particular, we are interested in demonstrating the potential for hierarchical localization. Colloquially, the CEO of a company should be more likely to be active in important discussions than a lower-level manager, even if both have the same number of direct subordinates. In other words, local connectivity alone does not fully characterize the potential for localization. And we are then also interested in how different structural features interact and how different types of localization might combine and amplify each other. How much more robust is localization around a central team than it is around a central star or a peripheral clique?

To this end, we develop an analytical description of organizational interaction network structure based on clique size distribution, the external degree distribution of cliques (inter-clique connectivity), and the hierarchical structure of the clique network as described by a refined $k$-core decomposition called onion decomposition~\cite{hebert-dufresne2016multiscale}. We then describe nonlinear dynamics on these networks using an approximate master equation (AME) framework~\cite{hebert-dufresne2010propagation, st-onge2021social, st-onge2021master} which allows us to analytically demonstrate hierarchical localization and numerically investigate its interplay with other types of dynamical localization.

\section{Construction of the Clustered Onion Network Ensemble}

We look at networks of teams (higher order networks which can reduce to usual networks for teams of size one) and describe them analytically using the hierarchy and connectivity patterns between teams. The layered structure of the hierarchy is specified by the onion decomposition~\cite{hebert-dufresne2016multiscale}, a refined version of the $k$-core decomposition.
The onion decomposition follows the same pruning algorithm but keeps track not only of the cores in which a node is found, but also of the layers in which they are removed in the pruning process.
Starting with an entire network as the $0$-core, nodes of degree 0 are removed to define the zeroth layer of the onion decomposition and leave us with the $1$-core of the network.
We then remove nodes of degree 1 in layer 1, and we remove nodes that are now of degree 1 (if any) in layer 2, and so on until we are left with the $2$-core of the network.
The process then repeats, starting with nodes of degree $2$ and always increasing the layer count.
This structure can be captured in our model using the onion network ensemble (ONE)~\cite{hebert-dufresne2016multiscale} and described mathematically as a layered \& correlated configuration model \cite{allard2019percolation}.

Each node in the ONE is then assumed to represent a team of arbitrary size, modeled as a clique, and thereby producing a Clustered Onion Network Ensemble (CONE), see Fig.~\ref{fig:cartoon}. Therein, the structural role of each team is specified by a size $n$, a degree $k$ (connections with other teams) and an onion layer $\ell$ corresponding to core $c(\ell)$ in the $k$-core decomposition (team centrality). 
We denote $N(x)$, $N(x,y)$ and $N(x,y,z)$ the distributions of teams where $x$, $y$, and $z$ can represent their size $n$, degree $k$, or layer $\ell$ (and any combination thereof) over the 3-dimensional space $(n,k,\ell)$.
Links between teams are assigned uniformly at random between the members of the respective teams, but the teams themselves must respect the underlying onion structure as shown in Fig.~\ref{fig:cartoon}. Teams of size 1 are individual nodes, teams of size 2 are simple edges but can be defined as a team to avoid imposing a hierarchical relationship between the two nodes involved, teams of size 3 are triangles, and so on.

For the organization to respect a hierarchy specified as an onion decomposition, teams of degree $k$ in layer $\ell$ must have exactly $c(\ell)$ links to layers $\ell'\geq \ell$ if they are in the first layer of their core, and otherwise at least $c(\ell)+1$ links to layers $\ell'\geq \ell-1$ and at most $c(\ell)$ links to layers $\ell'\geq \ell$ \cite{hebert-dufresne2016multiscale}. 
One can think of these links as those remaining once a node is reached by the pruning process of the $k$-core decomposition.
In fact, we call degrees connecting layer $\ell$ to layer $\ell'<\ell-1$ green or peripheral stubs (and these do not contribute to their position in the $k$-core decomposition), degrees connecting them to layer $\ell'=\ell-1$ blue or anchor stubs (as they potentially anchor them to their layer and can contribute to their position of the $k$-core decomposition), and degrees connecting them to layer $\ell'\geq \ell$ red or central stubs (which always contribute to the position of the $k$-core decomposition). 
The color naming scheme for stub types is used for visualization purposes, but colors always map to a concrete structural role: green stubs connect to more peripheral neighbors, blue stubs anchor nodes to the previous layer, and red stubs connect towards the core and contribute to the onion centrality.
These different links across layers are specified by a joint node-type connection matrix $L(k,\ell,k',\ell')$ counting the number of stubs starting from nodes of connectivity $(\ell,k)$ and leading to nodes in $(\ell',k')$.
The entire structure of this model is represented in Fig.~\ref{fig:cartoon}, starting from the conceptualization of a network in teams and centrality to the layer structure of the CONE.

\section{Approximate Master Equations} 

We describe the spread of information (ideas, norms, etc.) as a susceptible-infectious-susceptible (SIS) contagion model using an approximate master equation (AME) framework. This model is not meant to capture any specific spreading mechanisms, but simply describe how infectious nodes can transmit to their susceptible neighbors at a given rate $\lambda$ and recover at a fixed unit rate to become susceptible again. In the context of organisational network, one can think of the SIS model as describing how up-to-date nodes (infectious) that possess all information relevant to the organization can update their neighbors who are behind or uninformed (transmission to susceptible neighbors) but also fall behind themselves (recovery). Regardless of the exact conceptualization, the SIS dynamics is a minimal tractable model that captures how information can emerge, spread, and remain active on a given network structure.

We follow the dynamics on a networked, hierarchical, organization of teams. We denote $C_{i\vert n,k,\ell}$ the fraction of these teams which have a given structural role (size $n$, connectivity $k$, and onion centrality $\ell$) and have $i$ nodes currently ``infectious'' or active. We use SIS dynamics with a general infection kernel $\lambda(i,n,k,\ell)$ that can depend on the number of infectious nodes in a team (e.g., social reinforcement if $\lambda \propto i$), or on the implicit authority of a node (e.g. structural features $k$ and $\ell$). This general process can then be followed through the following set of AMEs
\begin{align}
    \dot{C}_{i\vert n,k,\ell} = & -(n-i)\left[i\lambda(i,n,k,\ell)
    + S_{n,k,\ell}^\mathrm{r} + S_{n,k,\ell}^\mathrm{g} + S_{n,k,\ell}^\mathrm{b} \right]C_{i\vert n,k,\ell} - iC_{i\vert n,k,\ell} \nonumber \\
    & +(n-i+1)\left[(i-1)\lambda(i-1,n,k,\ell)
    + S_{n,k,\ell}^\mathrm{r} + S_{n,k,\ell}^\mathrm{g} + S_{n,k,\ell}^\mathrm{b}\right]C_{i-1\vert n,k,\ell} + (i+1)C_{i+1\vert n,k,\ell}
    \label{eq:dotC}
\end{align}
where the $S^\mathrm{x}_{n,k,\ell}$ are the mean-field quantities coupling different groups through stubs of a given color. 
On average, any stub of a node of degree $k$ and layer $\ell$ will be red with probability
\begin{subequations}
\label{eq:stub_prob}
\begin{equation}
p^\mathrm{r}_{k,\ell} = \dfrac{\sum_{k',\ell'\geq \ell} L(k,\ell,k',\ell')}{c(\ell)N(k,\ell)}
\end{equation}
Similarly, the remaining stubs will be green with probability
\begin{equation}
   p^\mathrm{g}_{k,\ell} = \dfrac{\sum_{k',\ell'< \ell-1} L(k,\ell,k',\ell')}{\left[k-c(\ell)\right]N(k,\ell)}
\end{equation}
\end{subequations}
With these frequencies of stubs of different colors, we can calculate the average degree of each color for a node with a given connectivity $(k,\ell)$,
\begin{subequations}
\label{eq:avg_degree}
\begin{align}
  \langle k^\mathrm{r} \rangle_{k,\ell} & = c(\ell) p_{k,\ell}^\mathrm{r} \\
  \langle k^\mathrm{g} \rangle_{k,\ell} & = 
  \left\{
  \begin{array}{lc}
    (k - c(\ell)) p_{k,\ell}^\mathrm{g} - \delta_{c(\ell), c(\ell-1)} [p_{k,\ell}^\mathrm{r}]^{c(\ell)} (k - c(\ell)) p_{k,\ell}^\mathrm{g} & \text{if } k - c(\ell) \leq 1 \\
    &\\
    (k - c(\ell)) p_{k,\ell}^\mathrm{g}                                                                  & \text{otherwise}
  \end{array}\right. \\
  \langle k^\mathrm{b} \rangle_{k,\ell} & =
  \left\{
  \begin{array}{lc}
    c(\ell) (1 - p_{k,\ell}^\mathrm{g}) + (k - c(\ell)) (1 - p_{k,\ell}^\mathrm{g}) \\
    \qquad\qquad +\ \delta_{c(\ell), c(\ell-1)} [p_{k,\ell}^\mathrm{r}]^{c(\ell)} (k - c(\ell)) p_{k,\ell}^\mathrm{g} & \text{if } k - c(\ell) \leq 1 \\
    &\\
    c(\ell) (1 - p_{k,\ell}^\mathrm{r}) + (k - c(\ell)) (1 - p_{k,\ell}^\mathrm{g})                                                                                      & \text{otherwise.}
  \end{array}
  \right.
\end{align}
\end{subequations}
Altogether, this colored stub matching scheme allows us to evaluate the mean-field couplings, one color of stub at a time. First, for green stubs, we write
\begin{subequations}
\label{eq:mean_field_S}
\begin{equation}
    S_{n,k,\ell}^\mathrm{g} = \frac{\langle k^\mathrm{g}\rangle_{k,\ell}}{n} \sum_{\ell'<\ell-1,n',k',i'} \lambda(1,n',k',\ell') \dfrac{L(k,\ell,k',\ell')}{{\sum_{k'',\ell''<\ell-1}L(k,\ell,k'',\ell'')}} \frac{i'}{n'} C_{i'\vert n',k',\ell'} \; ,
\end{equation}
which is constructed as follows: Expected number of green stubs times the infection kernel of potential neighbors summed over all possible layers and cliques for these neighbors (biased by the layer-layer connection matrix $L$ and their expected number of red stubs). For blue stubs, we write
\begin{equation}
        S_{n,k,\ell}^\mathrm{b} = \frac{\langle k^\mathrm{b} \rangle_{k,\ell}}{n} \sum_{\ell'=\ell-1,n',k',i'} \lambda(1,n',k',\ell') \dfrac{L(k,\ell,k',\ell')}{{\sum_{k'', \ell''=\ell-1}L(k,\ell,k'',\ell'')}} \frac{i'}{n'} C_{i'\vert n',k',\ell'} \; ,
\end{equation}
and for red stubs,
\begin{align}
    S_{n,k,\ell}^\mathrm{r} =  \frac{\langle k^\mathrm{r} \rangle_{k,\ell}}{n} \sum_{\ell'\geq\ell,n',k',i'}  \lambda(1,n',k',\ell')\dfrac{L(k,\ell,k',\ell')}{\sum_{k'',\ell''\geq \ell} L(k,\ell,k'',\ell'')}\frac{i'}{n'}C_{i'\vert n',k',\ell'}
    \; .
\end{align}
\end{subequations}
Equations~\eqref{eq:dotC}--\eqref{eq:mean_field_S} provide a closed system of equations to follow generalized SIS dynamics on the CONE. This model describes network based on a unique combination of degree heterogeneity, clique structure and centrality pattern. However, the model does not capture correlations between neighboring nodes and teams; destroyed by the annealed structure of the master equations which averages over all nodes with a given structural description $(n,k,\ell)$. 
That being said, the description remains straightforward such that similar systems of equations could be written to follow other types of social dynamics. For example a voter model \cite{redner2019realityinspired} could be modeled by having recovery terms proportional to susceptible nodes in the team and introducing new mean-field quantities for the expected number of susceptible neighbors reached through the different types of stubs.

\section{Hierarchical localization}

\begin{figure}
    \centering
    \includegraphics[width=\linewidth]{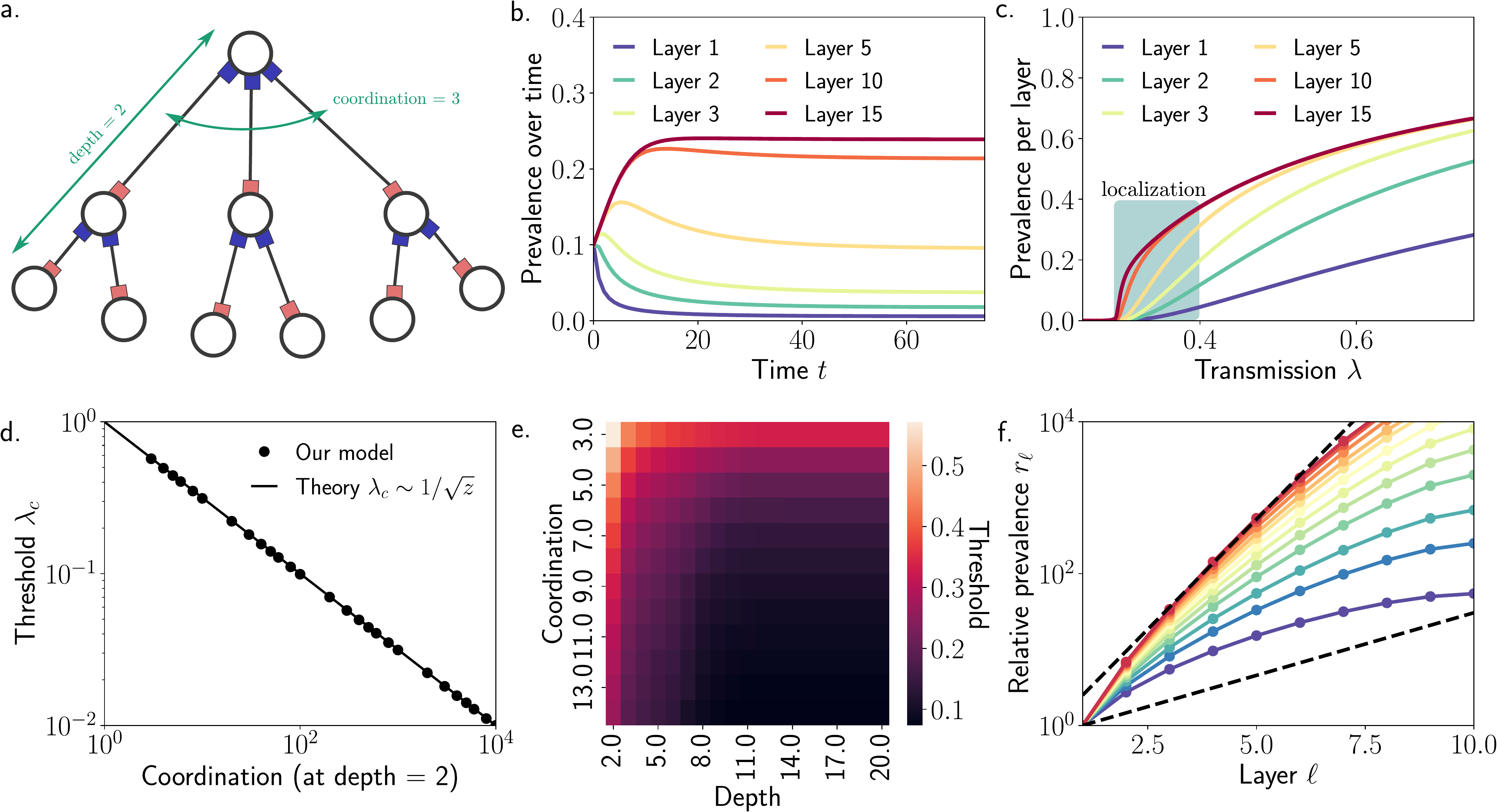}
    \caption{\textbf{Localization on a Cayley tree.} \textbf{(a.)} Parameterization of finite Cayley trees, and their stub structure in the CONE. \textbf{(b.)} Prevalence of contagion over time for individual layers on a Cayley tree of 20 layers with coordination number 4 and a transmission rate $\lambda = 0.33$. This shows how the dynamics can be supercritical in some layers with minimal spillover in others.  \textbf{(c.)} We fully explore the steady-state prevalence per layer (i.e., as time goes to infinity) in the same Cayley tree of 20 layers with coordination number 4. We see how the global threshold $\lambda_c \simeq 0.28$ is driven by the central layers only. \textbf{(d.)} Global activation threshold $\lambda_c$ of a star motif, i.e. a tree of depth $\ell_{\textrm{max}} = 2$, with varying coordination number (or degree). \textbf{(e.)} Global activation threshold $\lambda_c$ on the CONE from Cayley trees with varying depth and coordination number. \textbf{(f.)} We demonstrate hierarchical localization through the relative prevalence $r_{\ell}$ of layer $\ell$ versus layer 1, evaluated at the global activation threshold $\lambda_c$ for Cayley trees of depth 10 and coordination number between 3 (bottom blue curve) and 15 (top red curve). The two dashed lines show the exponential factor from Eq.~(\ref{eq:exponential}) without the additional $\ell$ dependencies, parameterized to match the bottom and top curves and therefore matching their main behavior observed in the middle layers where additional factors are less important.}
    \label{fig:starXlayer}
\end{figure}

We confirm that the CONE can capture dynamical localization around central nodes, at both the microscopic (degree) and macroscopic (onion centrality) levels in Fig.~\ref{fig:starXlayer}. As a first experiment, we simulate the SIS dynamics using Eqs.~\eqref{eq:dotC}--\eqref{eq:mean_field_S} on regular Cayley trees parameterized by their depth (denoted $\ell_{\textrm{max}})$) and coordination number (the degree of non-leaf nodes, denoted $z$) as shown in Fig.~\ref{fig:starXlayer}(a). Note that this section completely ignores both the team structure of the CONE, i.e. we fix $n=1$, as well as the complex contagion mechanisms, i.e. we fix $\lambda(i,n,k,\ell) = \lambda$.

Figure \ref{fig:starXlayer}(b) shows the temporal evolution of the activity level (i.e., the \textit{prevalence} or expected fraction of infectious nodes at a given layer) starting from a uniform initial condition in a tree with depth 20 and coordination number 4. While all layers $\ell > 1$ have the same connectivity, we see very different time series across different layers driven by differences in hierarchical position: With activity relaxing to very low levels in the peripheral layers while more central layers are orders of magnitude more active. Figure \ref{fig:starXlayer}(c) then shows the expected steady state prevalence (or infinite time limit) in the different layers of the tree as we vary the transmission rate. These results provide a first glimpse of hierarchical localization since the activation of the system around the global $\lambda_c = 0.3$ is driven only by some of the core layers of the tree. Conversely, layer 3 for example appears to activate only around 0.35 if we use the maximum derivative of its prevalence curve with respect to $\lambda$ as a proxy for its peak susceptibility \cite{hebert-dufresne2019smeared}.

Figure \ref{fig:starXlayer}(d) shows the value of the global threshold at a fixed depth of two and therefore looking at star motifs with different coordination number or degree. The CONE captures the previous key theoretical result from spectral analysis~\cite{goltsev2012localization}: The threshold for microscopic localization or activation around a star motif goes as the inverse of the square root of the star degree. More generally, Fig.~\ref{fig:starXlayer}(e) shows the global threshold $\lambda_c$ as a function of the depth and coordination number of the trees; illustrating the relative symmetry of the system and how adding hierarchical layers and local degree is somewhat equivalent. 

We can investigate these previous results analytically by considering a Cayley tree of depth $L$ with coordination number $z = d+1$ where $d$ is the number of descendants or children of non-leaf nodes.
Given the one-to-one mapping between degree and layer, we can directly write the CONE equation for the fraction of infected nodes $I_\ell$ in the layer $\ell$ as
\begin{align}
    \dot{I_\ell} = -I_l + (1-I_\ell) \left [ \beta (d + \delta_{\ell,\ell_\mathrm{max}}) I_{\ell-1} + \beta I_{\ell+1} \right ]\;,
\end{align}
where it is implicit that $I_0 = I_{\ell_\mathrm{max}+1} = 0$.
In the stationary state and near the epidemic threshold (where $I_\ell^* \to 0$ for all $\ell$, see Appendix), we can write the following recursive equation 
\begin{align}
    I_\ell^* = \frac{1}{\beta} I_{\ell-1}^* - d I_{\ell-2}^* \;,
\end{align}
which is valid for $\ell \in \{3,L\}$, where we define $L \equiv \ell_\mathrm{max}-1$.
We also have the boundary conditions $I_2^* = I_1^*/\beta$ and $I_{\ell_\mathrm{max}}^* = \beta (d+1) I_L^*$.
Since $I_\ell \to 0$, we are in fact interested in the relative quantity $r_\ell \equiv I_\ell^*/I_1^*$, which also respect the recursion
\begin{align}
    \label{eq:recursion_cayley}
   r_\ell =  \frac{1}{\beta} r_{\ell-1} - d r_{\ell-2} \;,
\end{align}
with boundary conditions $r_1 = 1$, $r_2 = 1/\beta$ and $r_{\ell_\mathrm{max}} = \beta (d+1) r_L$.
We define following generating function
\begin{align}
    G(x) = \sum_{\ell = 1}^\infty r_\ell x^\ell \;.
\end{align}
but are only interested in its first $L$ terms.
By multiplying Eq.~\eqref{eq:recursion_cayley} by $x^\ell$ and summing from $\ell = 3$ to $\ell \to \infty$\footnote{We are extending the validity of the recursion beyond $\ell = L$ without consequences as we will then only extract the first $L$ terms of the generating function.}, we obtain
\begin{align}
    G(x) - r_1 x-r_2x^2 = \frac{x}{\beta} \left [ G(x) - r_1x \right] - dx^2 G(x)  \;.
\end{align}
Rearranging the terms and using the boundary conditions to simplify the equation, we obtain
\begin{align}
    G(x) = \frac{x}{1 - \frac{x}{\beta} + dx^2} = \frac{x}{d (x-x_+)(x-x_-)} \;,
\end{align}
where
\begin{align}
    x_\pm = \frac{1 \pm \Delta}{2 d \beta}\;,\quad \Delta = \sqrt{1-4\beta^2d}\;.
\end{align}
If $\Delta \neq 0$, we can use the following partial fraction decomposition
\begin{align}
    G(x) = \frac{\beta x}{\Delta} \left [ \frac{1}{x-x_+} - \frac{1}{x-x_-} \right]\;,
\end{align}
and develop both terms using their Maclaurin series, which results in
\begin{align}
    G(x) = \frac{\beta}{\Delta} \left [ \sum_{\ell = 1}^\infty \left ( \frac{x}{x_-}\right)^\ell - \sum_{\ell = 1}^\infty \left ( \frac{x}{x_+}\right)^\ell \right]\;.
\end{align}
Therefore, for all $\ell \in \{1,L\}$,
\begin{align}
    r_\ell = \frac{\beta}{\Delta} \left [ \left( \frac{1}{x_-}\right)^\ell + \left ( \frac{1}{x_+}\right)^\ell\right] \;,
\end{align}
which is rewritten as
\begin{align}
    r_\ell = \frac{\beta}{\Delta} \left ( \frac{1}{2\beta}\right)^\ell \left [ (1+\Delta)^\ell - (1-\Delta)^\ell \right] \;.
\end{align}

\begin{figure}
    \centering
    \includegraphics[width=0.8\linewidth]{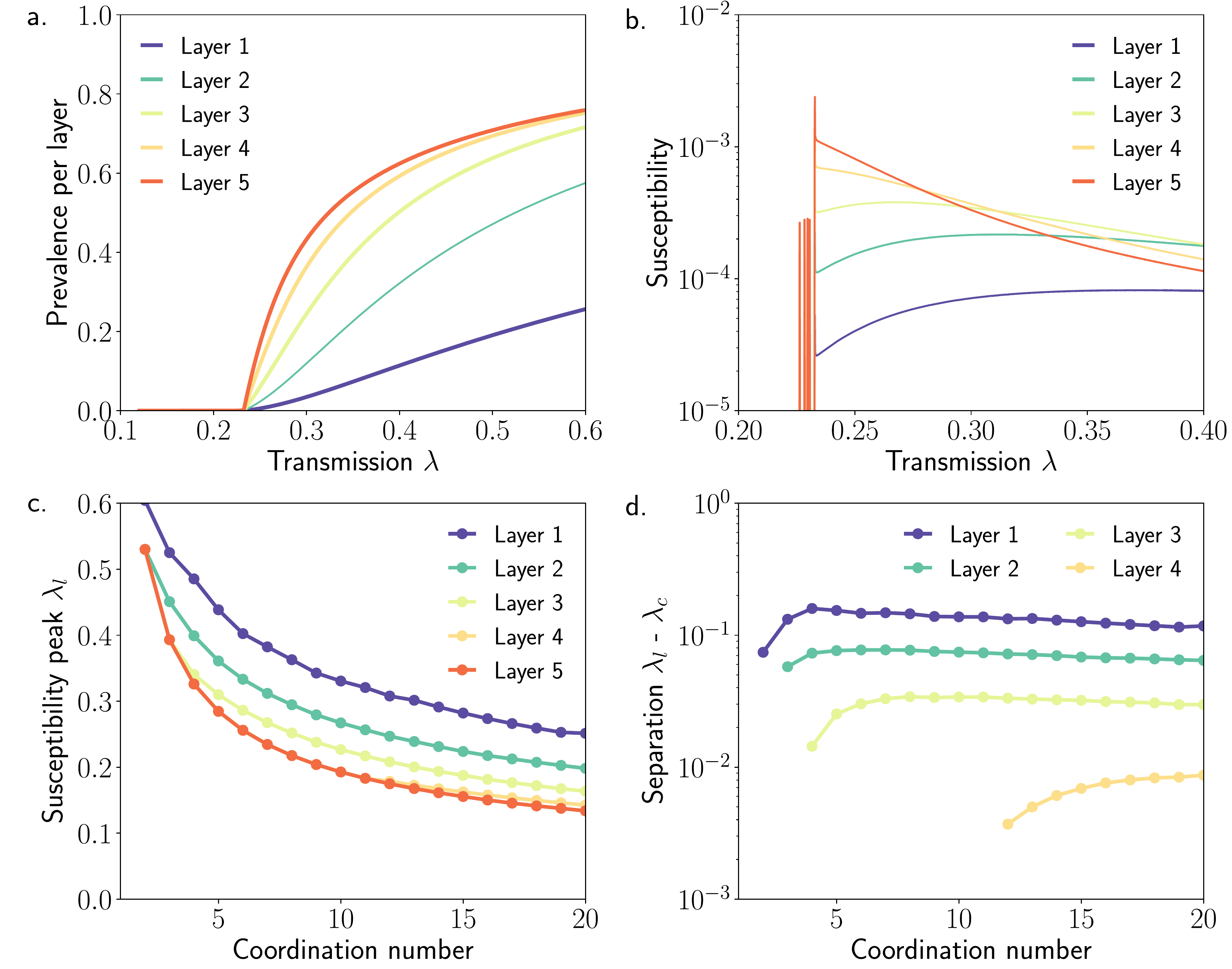}
    \caption{\textbf{Analysis of hierarchical localization.} \textbf{(a.)} Prevalence per layer in a Cayley tree of 5 layers with coordination number 7. \textbf{(b.)} Susceptibility, approximated by a numerical derivative over the prevalence, at each layer of the tree as a function of transmission rate $\lambda$. Peaks in susceptibility indicate important phase transitions: First as a global threshold to the diffusion process (with numerical noise), then, if any, a second peak can indicate a local activation threshold for a given layer. In this example, the dynamics are initially localized around layer 5 and 4, then grow to layer 3 as it activates around $\lambda = 0.25$, then to layer 2 around $\lambda = 0.3$ and finally to layer 1 close to $\lambda = 0.4$ \textbf{(c.)} Activation threshold per layer as we vary the coordination number of the Cayley tree with 5 layers. Hierarchical localization around specific layers requires a sufficiently large coordination number. For example, the dynamics can localize exclusively in layers 4 and 5 only if the coordination number is at least 4, and localized solely on the root only if the coordination number is at least 12. Any contagion with transmission rate between the two most extreme curves can therefore exhibit some level of hierarchical localization; whereas contagions fail to survive under the lower bound and collective delocalized dynamics occur above the upper bound. \textbf{(d.)} Plot of the differences between the local susceptibility peak $\lambda_{\ell}$ and the global threshold $\lambda_{c}$, showing the phase transitions behind the emergence of specific localization patterns.}
    \label{fig:hierarchical}
\end{figure} 

Note that in the limit $\Delta \to 0$, the preceding equation is well-defined
\begin{align}
    \lim_{\Delta \to 0} r_\ell = 2\beta \ell \left ( \frac{1}{2\beta}\right)^\ell \;,
\end{align}
 even though the partial fraction decomposition should have been computed differently.
 If $\Delta$ is imaginary ($4\beta^2 d > 1$), which should be the case given that the threshold of any tree has the threshold of the root $\beta_c \sim 1/\sqrt{z}$ as an upper bound (see Appendix), we can rewrite the solution as
 \begin{align}
     r_\ell = \frac{2\beta}{||\Delta||} (2d\beta)^\ell \sin (\theta \ell) \;,\quad \theta = \arcsin\left( \frac{||\Delta||}{4\beta^2d}\right) \;.
     \label{eq:exponential}
 \end{align}
 
Above the global threshold of the network we expect $2d\beta > 1$, defining a regime close to the threshold where we find an exponential relationship between a node's depth in the network (layer $\ell$) and its level of activity: Hierarchical localization. The phenomenon is confirmed in the last panel of Fig.~\ref{fig:starXlayer}.

We further explore hierarchical localization numerically in Fig.~\ref{fig:hierarchical}. Borrowing tools from statistical physics and the study of phase transitions, we can use the derivative of local prevalence per layer as a proxy for the susceptibility (here used in the statistical mechanics sense: the response of a system to changes in parameters). This quantity gives us a tool to identify local activation and further study the localization phenomenon \cite{hebert-dufresne2019smeared}. Figure \ref{fig:hierarchical}(a) presents the prevalence diagram on a tree with only 5 layers. In panel (b), we then show the discrete derivative which should peak at phase transitions. Here, these curves show a global maximum at the global activation threshold of the system (a classic phase transition) and local maxima at the local activation thresholds of unique layers that do not participate in the global activation. Before a local peak of layer 2, for example, the dynamics are localized in layers at least greater than 2.

In Fig.~\ref{fig:hierarchical}(c) we automate the detection of susceptibility peaks and tune the coordination number of the 5-layer tree. Because of the interplay between microscopic and hierarchical localization, we find that certain localization patterns require a minimal number of layers and degree. For example, localizing the dynamics solely on the root of a 5-layer tree requires a coordination number of at least 12. These specific localization patterns can emerge discontinuously, as shown in Fig.~\ref{fig:hierarchical}(d). As we tune the structure of a network (through discrete changes because of the discrete nature of networks), we can create explosive localization patterns: New susceptibility peaks can appear far from the global activation threshold. All of our results show the need to consider dynamical localization as a multidimensional phenomenon with multiple causes at different scales, like local connectivity and global layer structure, all of which can interact in interesting ways.

\section{Exploration of multidimensional localization}

Localization around stars or hubs has been a very active topic of research for the last decade, and here we have illustrated how it is a subset of hierarchical localization around trees of depth 2. Other structural mechanisms for localization (e.g., cliques) can be thought of as independent dimensions of the localization phenomenon. Using simple \textit{in silico} experiments, we now investigate how hierarchical localization can be amplified or hampered by localization around other network features.

\begin{figure}
    \centering
    \includegraphics[width=0.8\linewidth]{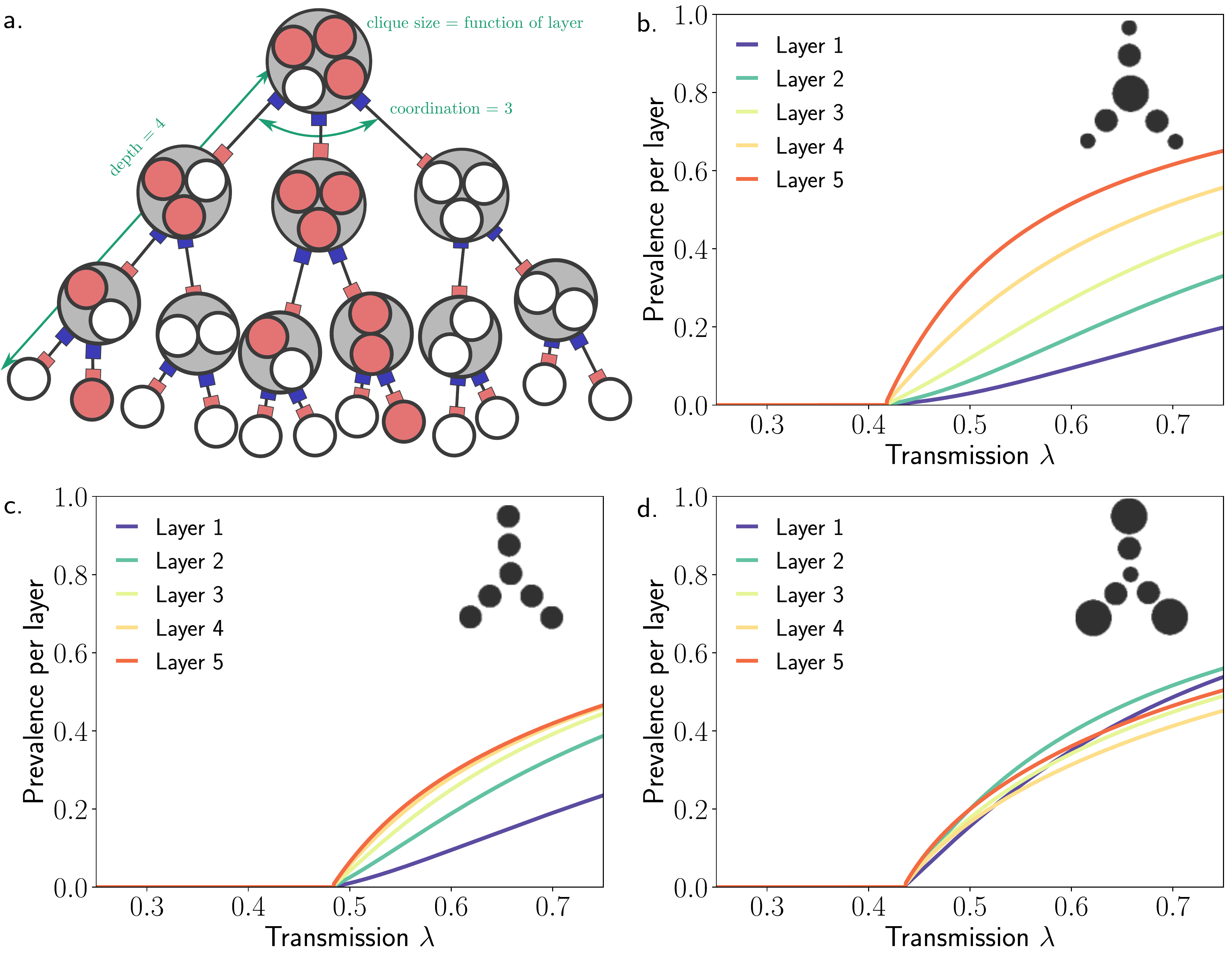}
    \caption{\textbf{Amplifying and cancelling localization on a tree of cliques.} \textbf{(a.)} Parameterization of finite trees of cliques where the tree structure is enforced by the stub structure of the CONE and the sizes of cliques are specified as a function of their layer. \textbf{(b.)} We reinforce localization by setting clique size $n$ equal to layer $\ell$: Teams at the top of the hierarchy are larger. We use depth $\ell_{\textrm{max}} = 5$ and coordination number $z=3$. \textbf{(c.)} We use the same tree and set all cliques size equal to 3. \textbf{(d.)} Again using the same tree, we delocalize the dynamics by setting clique size $n$ inversely proportional to layer: $n = \ell_{\textrm{max}} + 1 - \ell$ such that teams at the bottom of the hierarchy are larger.}
    \label{fig:cliques}
\end{figure}

First, we can use the clique formulation of the CONE to include elements of mesoscopic localization. This allows us to look at the interplay of macroscopic hierarchy and mesoscopic cliques by using the CONE to study dynamics on a tree of cliques (or a hierarchy of teams). Organizations tend to have a pyramid structures as a result of power centralization \cite{pennings1973measures}, which suggest a strong potential for hierarchical localization. However, in these pyramids, team size can be either (1) positively correlated with onion layers such that teams higher in the hierarchy are also larger, (2) negatively correlated with layers such that peripheral teams tend to be larger, or (3) uncorrelated such that all teams have an average size regardless of their position in the hierarchy. Note that correlations between team size and centrality changes the density of the resulting network, but with surprising impacts on the dynamics. We show these results in Fig.~\ref{fig:cliques}. Our results illustrate how cliques can amplify hierarchical localization through positive correlations, as we find a larger separation of susceptibility in scenario 1 compared to 3. Conversely, negative correlations can complete delocalize the dynamics, as seen in scenario 2. Interestingly, despite having more links in the network, scenario 3 delocalizes the dynamics at the price of having a higher global activation threshold than scenario 1.

\begin{figure}
    \centering
    \includegraphics[width=0.8\linewidth]{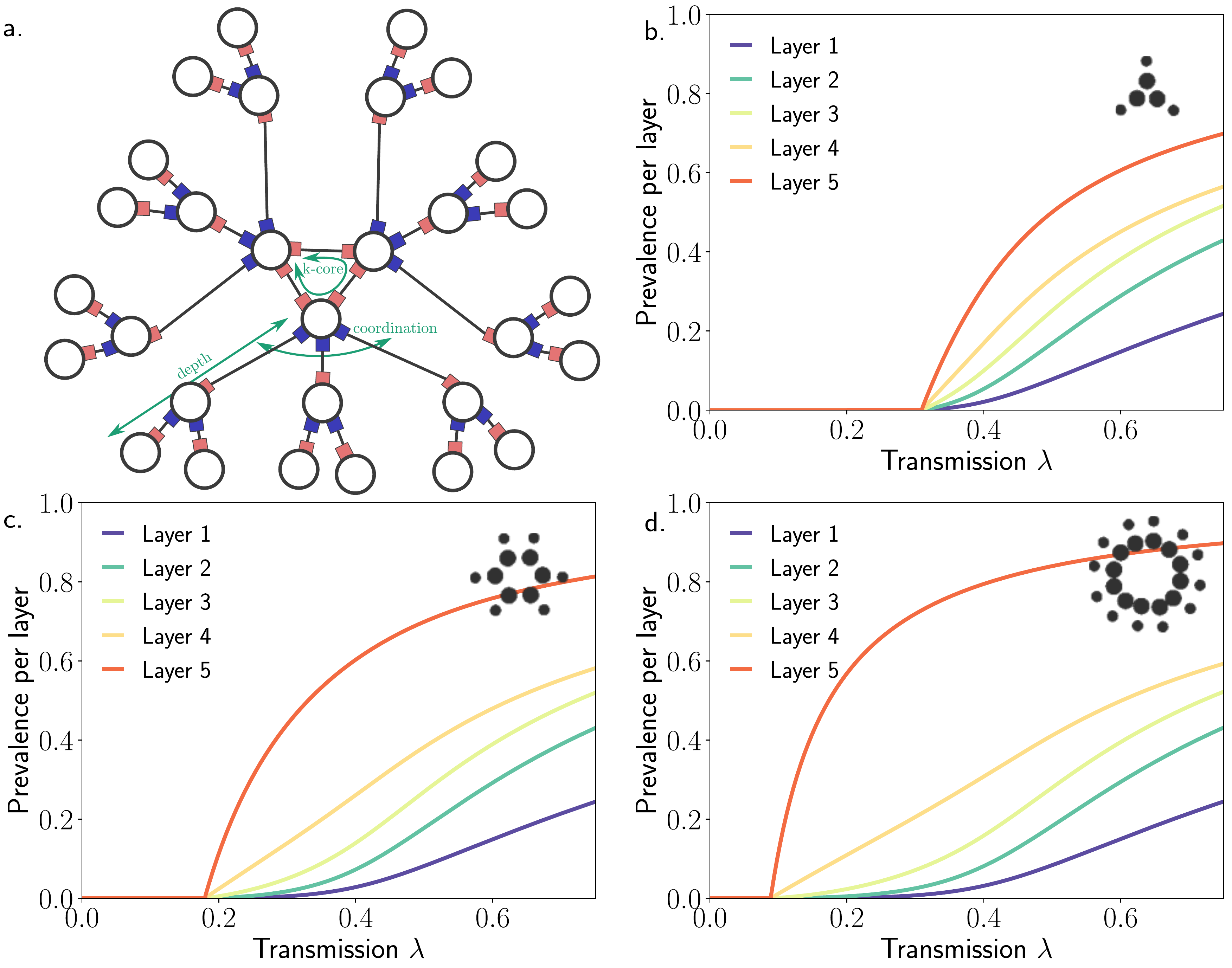}
    \caption{\textbf{Amplifying localization on a core with treelike periphery structure.} \textbf{(a.)} Parameterization of a uniform $k-core$ where all nodes are roots of finite trees, all enforced by the stub structure of the CONE. In panels \textbf{(b-d)}, we use a central group of 20 nodes that form a 2-, 5-, or 11-core respectively. All nodes in that core are the roots of their own finite trees of depth 5 and coordination number 3, resulting in a 5 layers of Onion Decomposition with a bimodal $k$-core structure (core vs periphery). As the $k$-core grows denser, network structure amplifies the localization phenomenon through a synergy between core localization around the roots and hierarchical localization in the periphery. Importantly, while the number of peripheral trees depends on the density of the core, the prevalence observed on each tree is almost independent of the prevalence in the core as inner layers provide a buffer zone (i.e., the blue curves barely change).}
    \label{fig:cores}
\end{figure}

\begin{figure}
    \centering
    \includegraphics[width=0.8\linewidth]{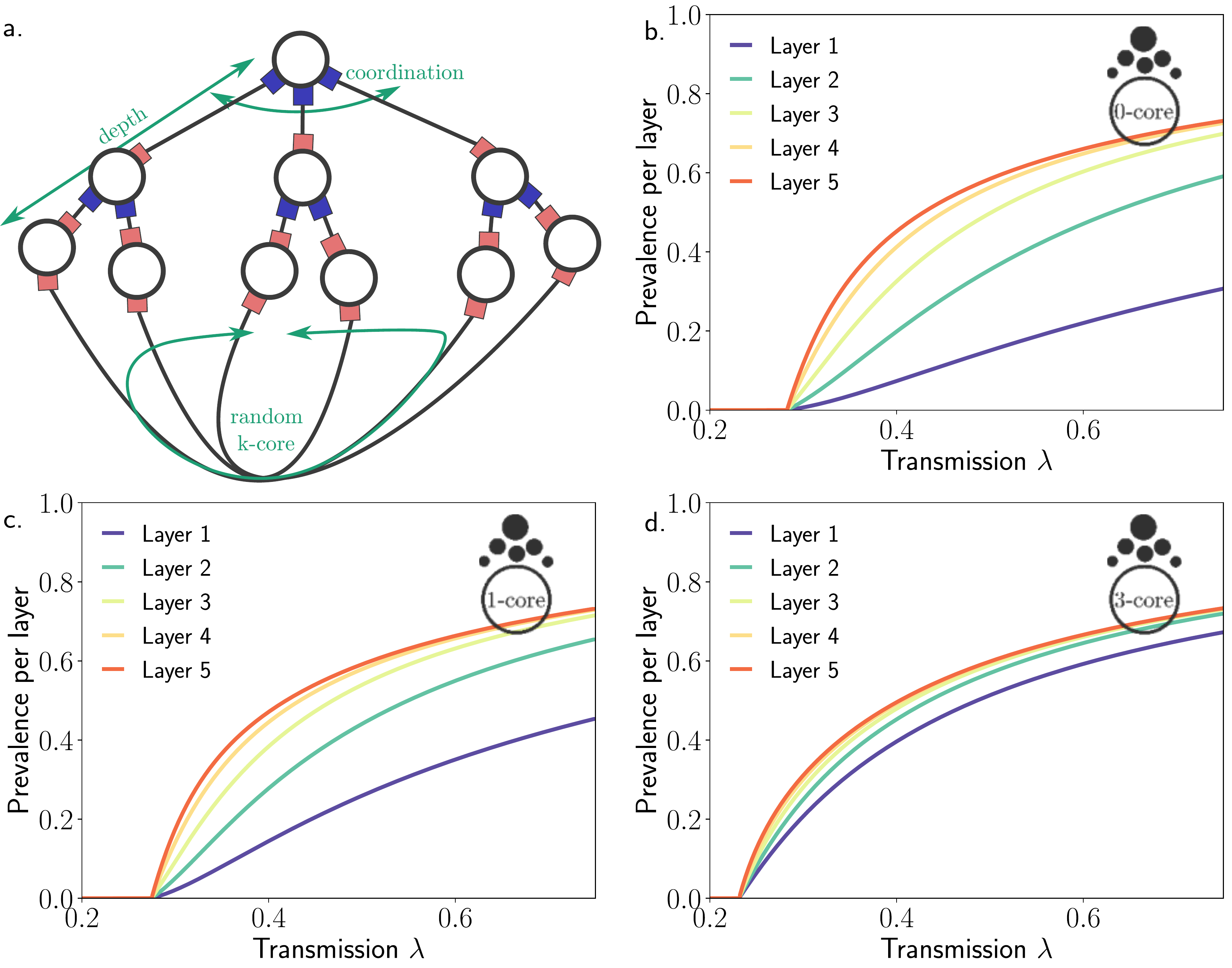}
    \caption{\textbf{Delocalizing dynamics on a tree through $k$-core structure in the periphery.} \textbf{(a.)} We start with a finite tree and connect all leaves as a uniform $k$-core. The resulting networks has a number of onion layers equal to the depth of the original tree and the root preserves its centrality but with all nodes in the $(k+1)$-core, as long as $k+1$ is smaller than the coordination number of the original tree. In panels \textbf{(b-d)}, we fix the depth and coordination number of the original tree to 5, and modify the peripheral $k$-core from 0 (control) to a 1-core and a 3-core. As the peripheral $k$-core grows denser, we lower the threshold of the whole system and also delocalize the dynamics away from the inner layers.}
    \label{fig:periphery}
\end{figure}

We then deviate from trees and leverage the ability of the CONE to produce networks with a fixed $k$-core structure. In Fig.~\ref{fig:cores}, we introduce a $k$-core at the top of the hierarchy. This of course amplifies the localization phenomenon by allowing the $k$-core of root nodes to self-activate. Interestingly, while this self-activation lowers the global threshold and spills over into the penultimate onion layers, it barely affects the lower layers of the hierarchy. The local activation threshold of nodes in layers 1 or remains essentially unchanged as we grow the inner $k$-core of the network. In Fig.~\ref{fig:periphery} we do the opposite and introduce a $k$-core among leaves in the periphery of the initial structure. This process is akin to team building exercise or socialization meant to disseminate information from one branch of an organization to another. We find that this process allows an organization to progressively delocalize the dynamics. Yet, unlike the results of Fig.~\ref{fig:cliques}, socialization across branches also simultaneously decreases the global activation threshold of the whole network. In other words, this type of structure allows easier and more homogeneous spread of information.

Finally, we briefly explore the generality of the CONE beyond network structure by using the $\lambda(i,n,k,\ell)$ function. This complex transmission rate can allow us model classic complex contagion where groups with more activity (i.e., higher $i$) diffusion information more effectively by setting $\lambda(i,n,k,\ell) \propto i$. This scenario has recently been well studied in the context of dynamics on higher-order networks \cite{Iacopini2019simplicial}. Such dynamics can lead to interesting transitions where activity emerges first in larger teams before diffusing to smaller ones. Using AMEs formulation in terms of group structure, we recently showed how this modifies classic notions of network influence \cite{st-onge2021influential}: In non-linear contagions, influence becomes a group or team property, rather than an individual property of network hubs.

One unique feature of the CONE is that we can now also model situations where teams or individuals with privileged network structure (larger size $n$, connectivity $k$ or centrality $\ell$) have access to more or less information, by setting for example $\lambda(i,n,k,\ell) \propto \ell^\gamma$ as we do in Fig.~\ref{fig:complex_l}. (Note that our analysis presented in Appendix can be generalized to this type of complex contagion.) When $\gamma$ is positive, central nodes receive more power and amplify the localization: We can saturate the dynamics around central layers while the periphery remains virtually inactive. More interestingly, we can also localize the dynamics around central layers by opposing power and network positions when $\gamma$ is negative. In this case, the periphery might perceive a higher transmission rate but remains at a structural disadvantage while the root sees a lower transmission rate despite its advantageous network position. In between, certain layers can strike the right balance and outperform both the center and periphery of the hierarchy. While weakly negative $\gamma$ values hinder localization [Fig.~\ref{fig:complex_l}(a.)], for sufficiently negative $\gamma$ [Fig.~\ref{fig:complex_l}(a.)] we find a phase transition where activity emerges in a strongly localized manner within these balanced central layers, before weakly diffusing asymmetrically towards the periphery and root. The most influential nodes are thus those that have the right balance of network centrality and dynamical activity. This nontrivial interplay of structure and dynamics define the \textit{influential layers} of the network, i.e., those able to sustain the dynamics on their own.

\begin{figure}
    \centering
    \includegraphics[width=0.8\linewidth]{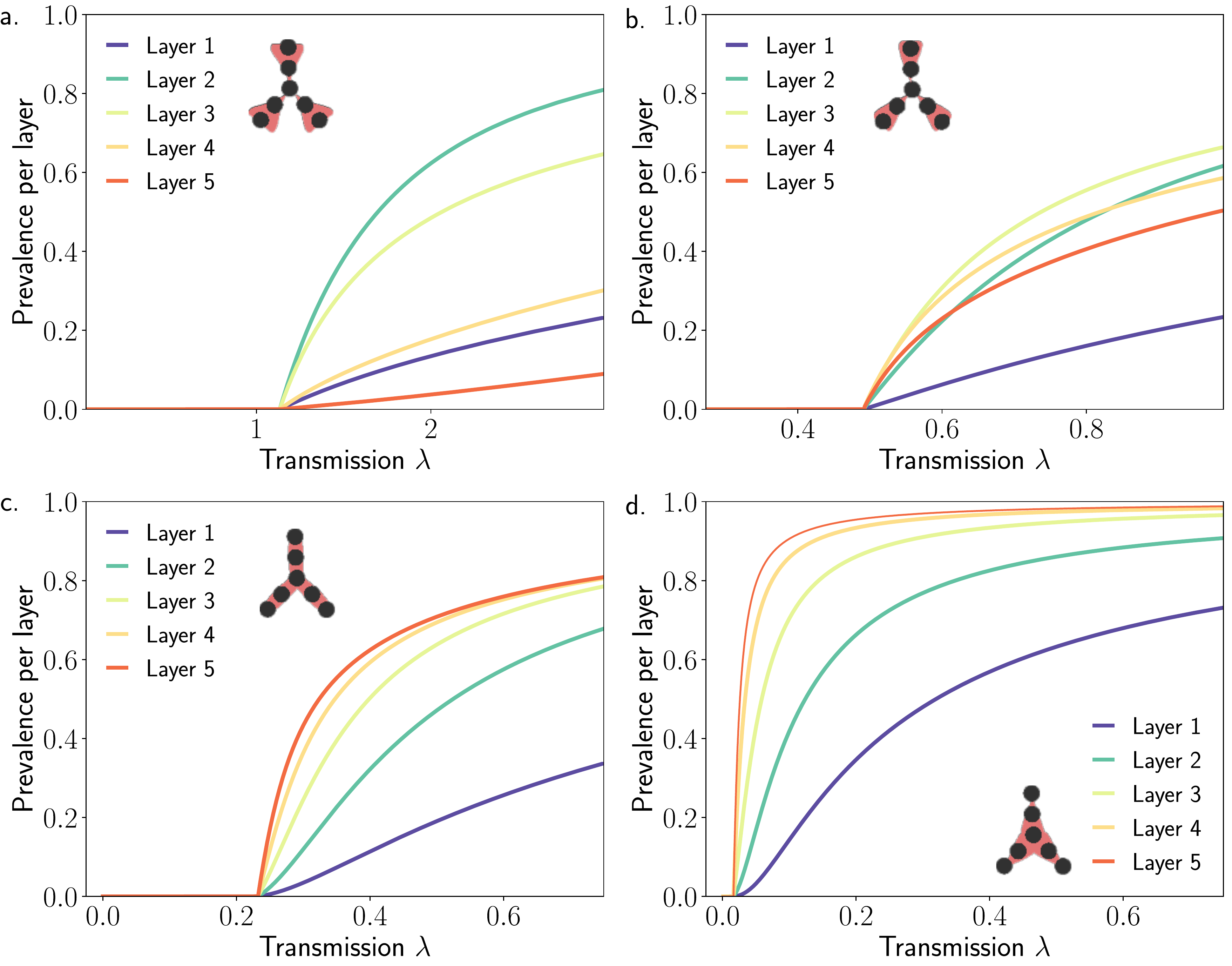}
    \caption{\textbf{Modifying hierarchical localization through a complex transmission function.} We use a finite tree of 5 layers with coordination number 7, as in Fig.~\ref{fig:hierarchical} and implement a transmission function that varies with layer: $\lambda(l) \equiv \lambda \times \ell^\gamma$. This function exogeneously models the fact that certain position in the hierarchical organization might have more power and or sensitivity to the contagion process. \textbf{(a.-b.)} We amplify the transmission rate in the periphery with $\gamma$ equal to -3.0 and -1.0, respectively. \textbf{(c.)} Classic contagion case with $\gamma = 0$. \textbf{(d.)} We amplify the power of core layers with $\gamma = 2.0$, lowering the global threshold and increase the extent of hierarchical localization.}
    \label{fig:complex_l}
\end{figure}

\section{Conclusion}

In this paper, we have made both concrete and forward-looking steps towards a better theoretical understanding of how dynamical processes unfold in organizational interaction networks. First, we provided a mathematical demonstration of hierarchical localization; showing how key layers of centrality can potentially support a dynamical process on their own. Second, we explored the multidimensional interplay of different localization mechanisms by varying degree distributions, centrality patterns, team sizes, core structure, and transmission functions. Through a series of experiments, we have highlighted how carefully designing organizational networks can either amplify or hinder dynamical localization.

This exploratory analysis paves the way for future work on multiple fronts. Future theoretical studies could attempt to optimize network structure to either enhance or limit siloing of information as desired by a given organization. More applied work could leverage the CONE to quantify the localization potential of real organizational structures through formal organizational charts, network surveys, or email archives. This could be done, for example, by measuring the standard deviation around the peak susceptibility of different teams based on their size, connectivity and centrality in the organization; i.e., the standard deviation over the values shown for a given network in Fig.~\ref{fig:hierarchical}(c). Using peaks of susceptibility means that localization and siloing potential are not measured against a specific dynamical process with fixed parameters, but simply against a spreading mechanism regardless of parameter values. This might provide a parameter-agnostic tool for the comparison of organizational networks.

Finally, our work contributes to important questions in organizational research: How do structures affect organizations' abilities to share different kinds of information? And, as others also ask\cite{majchrzak2011transcending, szulanski2016stickiness, edmondson2018crossboundary}, how do network properties moderate the role of factors like process, context, information content, and identity? Moreover, while researchers increasingly understand \textit{how} and \textit{why} knowledge embeds in organizational units, organizational scholarship still faces critical challenges. How can we limit the spread of stress-related disorders\cite{kensbock2021mental} without hindering the dissemination of knowledge? How can we prevent the diffusion of maladaptive behaviors, like corruption \cite{ferrali2020crime}, while promoting inclusive practices \cite{melusoInPressmasculinity} through socialization? The answer could lie in the complex interplay of information dynamics and network structure; by amplifying the transmission of certain types of information in key regions of an interaction network, one can intentionally localize or delocalize their spread. Our work introduced a network scientific technique for exploring solutions to important questions through greater understanding of localization within subsets of an organization, and of diffusion throughout organizational structures.

\subsection*{Acknowledgements}

The authors thank Jean-Gabriel Young for discussions and Google Open Source for support under the Open-Source Complex Ecosystems And Networks (OCEAN) project. 
L.H.-D. is also supported by the National Science Foundation Grant No. DMS-1829826 and A.A. by the Sentinelle Nord initiative of the Canada First Research Excellence Fund and the Natural Sciences and Engineering Research Council of Canada (project 2019-05183).
Any opinions, findings, and conclusions or recommendations expressed in this material are those of the authors and do not necessarily reflect the views of any of the funders.


\section*{Appendix: Epidemic threshold for the modified Cayley tree}

Let us consider a modified Cayley tree where the root node is of degree $d$ and each descendent of this node are of degree $d+1$, except for the leaves which are of degree $1$. This is also called a perfect $d-$ary tree. The CONE equations are then
\begin{align}
    \dot{I}_\ell = -I_\ell + (1-I_\ell) \left [ \beta d I_{\ell-1} + \beta I_{\ell+1}\right] \;.
\end{align}
Near the epidemic threshold, we can linearize the preceding equation and write it under matrix form
\begin{align}
    \frac{\mathrm{d} \boldsymbol{I}}{\mathrm{d} t} = \boldsymbol{M} \boldsymbol{I} \;,
\end{align}
where
\begin{align}
    \boldsymbol{M} = \begin{bmatrix} 
    -1 & \beta & 0 & \dots & 0 \\
    \beta d & -1 & \beta & & \\
    0 &  \beta d   & \ddots & \ddots & \\
    \vdots & & \ddots & \ddots & \beta \\
    0 & & & \beta d & -1 
    \end{bmatrix}
\end{align}
is a tridiagonal Toeplitz matrix. The eigenvalues $(\phi_j)_{j = 1}^{\ell_\mathrm{max}}$ of $\boldsymbol{M}$ are given by~\cite{noschese2013}
\begin{align}
    \phi_j = -1 + 2 \beta \sqrt{d}\cos \left ( \frac{j \pi}{\ell_{\mathrm{max}} + 1} \right) \;.
\end{align}
The critical point is reached when the largest eigenvalue
\begin{align}
    \Phi = -1 + 2\beta \sqrt{d} \cos \left ( \frac{ \pi}{\ell_{\mathrm{max}} + 1} \right) \;
\end{align}
equals zero. Therefore, the epidemic threshold is
\begin{align}
    \beta_\mathrm{c} = \frac{1}{2 \sqrt{d} \cos \left ( \frac{ \pi}{\ell_{\mathrm{max}} + 1} \right) } \;.
\end{align}
For $\ell_\mathrm{max}$ sufficiently large, we find $\beta_\mathrm{c} \simeq 1/(2\sqrt{d})$. Since the degree of the root node for the standard Cayley tree is larger, we expect the epidemic threshold of the Cayley tree to be upper-bounded by the expression above, i.e., $\beta_\mathrm{c} \lesssim 1/(2\sqrt{d})$.


\begin{thebibliography}{10}
\urlstyle{rm}
\expandafter\ifx\csname url\endcsname\relax
  \def\url#1{\texttt{#1}}\fi
\expandafter\ifx\csname urlprefix\endcsname\relax\def\urlprefix{URL }\fi
\expandafter\ifx\csname doiprefix\endcsname\relax\def\doiprefix{DOI: }\fi
\providecommand{\bibinfo}[2]{#2}
\providecommand{\eprint}[2][]{\url{#2}}

\bibitem{katz1963traditions}
\bibinfo{author}{Katz, E.}, \bibinfo{author}{Levin, M.~L.} \&
  \bibinfo{author}{Hamilton, H.}
\newblock \bibinfo{journal}{\bibinfo{title}{Traditions of research on the
  diffusion of innovation}}.
\newblock {\emph{\JournalTitle{American Sociological Review}}}
  \bibinfo{pages}{237--252} (\bibinfo{year}{1963}).

\bibitem{clement2018evolution}
\bibinfo{author}{Clement, J.} \& \bibinfo{author}{Puranam, P.}
\newblock \bibinfo{journal}{\bibinfo{title}{Searching for structure: formal
  organization design as a guide to network evolution}}.
\newblock {\emph{\JournalTitle{Management Science}}}
  \textbf{\bibinfo{volume}{64}}, \doiprefix\url{10.1287/mnsc.2017.2807}
  (\bibinfo{year}{2018}).

\bibitem{bento2020organizational}
\bibinfo{author}{Bento, F.}, \bibinfo{author}{Tagliabue, M.} \&
  \bibinfo{author}{Lorenzo, F.}
\newblock \bibinfo{journal}{\bibinfo{title}{Organizational silos: a scoping
  review informed by a behavioral perspective on systems and networks}}.
\newblock {\emph{\JournalTitle{Societies}}} \textbf{\bibinfo{volume}{10}},
  \bibinfo{pages}{56} (\bibinfo{year}{2020}).

\bibitem{ahuja1999network}
\bibinfo{author}{Ahuja, M.~K.} \& \bibinfo{author}{Carley, K.~M.}
\newblock \bibinfo{journal}{\bibinfo{title}{Network structure in virtual
  organizations}}.
\newblock {\emph{\JournalTitle{Organization Science}}}
  \textbf{\bibinfo{volume}{10}}, \bibinfo{pages}{741--757}
  (\bibinfo{year}{1999}).

\bibitem{long2007social}
\bibinfo{author}{Long, Y.} \& \bibinfo{author}{Siau, K.}
\newblock \bibinfo{journal}{\bibinfo{title}{Social network structures in open
  source software development teams}}.
\newblock {\emph{\JournalTitle{Journal of Database Management (JDM)}}}
  \textbf{\bibinfo{volume}{18}}, \bibinfo{pages}{25--40}
  (\bibinfo{year}{2007}).

\bibitem{hinds2008social}
\bibinfo{author}{Hinds, D.} \& \bibinfo{author}{Lee, R.~M.}
\newblock \bibinfo{title}{Social network structure as a critical success
  condition for virtual communities}.
\newblock In \emph{\bibinfo{booktitle}{Proceedings of the 41st Annual Hawaii
  International Conference on System Sciences (HICSS 2008)}},
  \bibinfo{pages}{323--323} (\bibinfo{organization}{IEEE},
  \bibinfo{year}{2008}).

\bibitem{evans2018opinion}
\bibinfo{author}{Evans, T.} \& \bibinfo{author}{Fu, F.}
\newblock \bibinfo{journal}{\bibinfo{title}{Opinion formation on dynamic
  networks: identifying conditions for the emergence of partisan echo
  chambers}}.
\newblock {\emph{\JournalTitle{Royal Society Open Science}}}
  \textbf{\bibinfo{volume}{5}}, \bibinfo{pages}{181122} (\bibinfo{year}{2018}).

\bibitem{hebert2017strategic}
\bibinfo{author}{H{\'e}bert-Dufresne, L.}, \bibinfo{author}{Allard, A.},
  \bibinfo{author}{No{\"e}l, P.-A.}, \bibinfo{author}{Young, J.-G.} \&
  \bibinfo{author}{Libby, E.}
\newblock \bibinfo{journal}{\bibinfo{title}{Strategic tradeoffs in competitor
  dynamics on adaptive networks}}.
\newblock {\emph{\JournalTitle{Scientific Reports}}}
  \textbf{\bibinfo{volume}{7}}, \bibinfo{pages}{1--11} (\bibinfo{year}{2017}).

\bibitem{long2013bridges}
\bibinfo{author}{Long, J.~C.}, \bibinfo{author}{Cunningham, F.~C.} \&
  \bibinfo{author}{Braithwaite, J.}
\newblock \bibinfo{journal}{\bibinfo{title}{Bridges, brokers and boundary
  spanners in collaborative networks: a systematic review}}.
\newblock {\emph{\JournalTitle{BMC Health Services Research}}}
  \textbf{\bibinfo{volume}{13}}, \bibinfo{pages}{1--13} (\bibinfo{year}{2013}).

\bibitem{melusoInPressmasculinity}
\bibinfo{author}{Meluso, J.}, \bibinfo{author}{Bagrow, J.},
  \bibinfo{author}{{H{\'e}bert-Dufresne}, L.} \& \bibinfo{author}{Razzante, R.}
\newblock \bibinfo{title}{Masculinity contest cultures and inclusive cultures:
  insights from an agent-based model of organizational socialization and
  promotion}.
\newblock In \bibinfo{editor}{King, E.~B.}, \bibinfo{editor}{Roberson, Q.~M.}
  \& \bibinfo{editor}{Hebl, M.} (eds.) \emph{\bibinfo{booktitle}{Research on
  social issues in management: the future of diversity \& inclusion}},
  vol.~\bibinfo{volume}{3} (\bibinfo{publisher}{Information Age Publishing},
  \bibinfo{year}{In Press}).

\bibitem{mitra2017spread}
\bibinfo{author}{Mitra, T.}, \bibinfo{author}{Muller, M.},
  \bibinfo{author}{Shami, N.~S.}, \bibinfo{author}{Golestani, A.} \&
  \bibinfo{author}{Masli, M.}
\newblock \bibinfo{journal}{\bibinfo{title}{Spread of employee engagement in a
  large organizational network: a longitudinal analysis}}.
\newblock {\emph{\JournalTitle{Proceedings of the ACM on Human-Computer
  Interaction}}} \textbf{\bibinfo{volume}{1}}, \bibinfo{pages}{1--20}
  (\bibinfo{year}{2017}).

\bibitem{granovetter1973weakties}
\bibinfo{author}{Granovetter, M.~S.}
\newblock \bibinfo{journal}{\bibinfo{title}{The strength of weak ties}}.
\newblock {\emph{\JournalTitle{American Journal of Sociology}}}
  \textbf{\bibinfo{volume}{78}}, \bibinfo{pages}{1360--1380}
  (\bibinfo{year}{1973}).

\bibitem{macy1990learning}
\bibinfo{author}{Macy, M.~W.}
\newblock \bibinfo{journal}{\bibinfo{title}{Learning {{Theory}} and the
  {{Logic}} of {{Critical Mass}}}}.
\newblock {\emph{\JournalTitle{American Sociological Review}}}
  \textbf{\bibinfo{volume}{55}}, \bibinfo{pages}{809--826},
  \doiprefix\url{10.2307/2095747} (\bibinfo{year}{1990}).

\bibitem{smith-crowe2014corruption}
\bibinfo{author}{{Smith-Crowe}, K.} \& \bibinfo{author}{Warren, D.~E.}
\newblock \bibinfo{journal}{\bibinfo{title}{The {{Emotion-Evoked Collective
  Corruption Model}}: {{The Role}} of {{Emotion}} in the {{Spread}} of
  {{Corruption Within Organizations}}}}.
\newblock {\emph{\JournalTitle{Organization Science}}}
  \textbf{\bibinfo{volume}{25}}, \bibinfo{pages}{1154--1171},
  \doiprefix\url{10.1287/orsc.2014.0896} (\bibinfo{year}{2014}).

\bibitem{watson2006knowledge}
\bibinfo{author}{Watson, S.} \& \bibinfo{author}{Hewett, K.}
\newblock \bibinfo{journal}{\bibinfo{title}{A {{Multi-Theoretical Model}} of
  {{Knowledge Transfer}} in {{Organizations}}: {{Determinants}} of {{Knowledge
  Contribution}} and {{Knowledge Reuse}}*}}.
\newblock {\emph{\JournalTitle{Journal of Management Studies}}}
  \textbf{\bibinfo{volume}{43}}, \bibinfo{pages}{141--173},
  \doiprefix\url{10.1111/j.1467-6486.2006.00586.x} (\bibinfo{year}{2006}).

\bibitem{ferrali2020crime}
\bibinfo{author}{Ferrali, R.}
\newblock \bibinfo{journal}{\bibinfo{title}{Partners in crime? {{Corruption}}
  as a criminal network}}.
\newblock {\emph{\JournalTitle{Games and Economic Behavior}}}
  \textbf{\bibinfo{volume}{124}}, \bibinfo{pages}{319--353},
  \doiprefix\url{10.1016/j.geb.2020.08.013} (\bibinfo{year}{2020}).

\bibitem{centola2007complex}
\bibinfo{author}{Centola, D.} \& \bibinfo{author}{Macy, M.}
\newblock \bibinfo{journal}{\bibinfo{title}{Complex contagions and the weakness
  of long ties}}.
\newblock {\emph{\JournalTitle{American Journal of Sociology}}}
  \textbf{\bibinfo{volume}{113}}, \bibinfo{pages}{702--734},
  \doiprefix\url{10.1086/521848} (\bibinfo{year}{2007}).

\bibitem{dougherty1992thoughtworlds}
\bibinfo{author}{Dougherty, D.}
\newblock \bibinfo{journal}{\bibinfo{title}{Interpretive barriers to successful
  product innovation in large firms}}.
\newblock {\emph{\JournalTitle{Organization science}}}
  \textbf{\bibinfo{volume}{3}}, \bibinfo{pages}{179--202}
  (\bibinfo{year}{1992}).

\bibitem{majchrzak2011transcending}
\bibinfo{author}{Majchrzak, A.}, \bibinfo{author}{More, P. H.~B.} \&
  \bibinfo{author}{Faraj, S.}
\newblock \bibinfo{journal}{\bibinfo{title}{Transcending knowledge differences
  in cross-functional teams}}.
\newblock {\emph{\JournalTitle{Organization Science}}}
  \textbf{\bibinfo{volume}{23}}, \bibinfo{pages}{951--970},
  \doiprefix\url{10.1287/orsc.1110.0677} (\bibinfo{year}{2011}).

\bibitem{wenger2000communities}
\bibinfo{author}{Wenger, E.}
\newblock \bibinfo{journal}{\bibinfo{title}{Communities of {{Practice}} and
  {{Social Learning Systems}}}}.
\newblock {\emph{\JournalTitle{Organization}}} \textbf{\bibinfo{volume}{7}},
  \bibinfo{pages}{225--246}, \doiprefix\url{10.1177/135050840072002}
  (\bibinfo{year}{2000}).

\bibitem{martin2014localization}
\bibinfo{author}{Martin, T.}, \bibinfo{author}{Zhang, X.} \&
  \bibinfo{author}{Newman, M.~E.}
\newblock \bibinfo{journal}{\bibinfo{title}{Localization and centrality in
  networks}}.
\newblock {\emph{\JournalTitle{Physical Review E}}}
  \textbf{\bibinfo{volume}{90}}, \bibinfo{pages}{052808}
  (\bibinfo{year}{2014}).

\bibitem{st-onge2021social}
\bibinfo{author}{{St-Onge}, G.}, \bibinfo{author}{Thibeault, V.},
  \bibinfo{author}{Allard, A.}, \bibinfo{author}{Dub{\'e}, L.~J.} \&
  \bibinfo{author}{{H{\'e}bert-Dufresne}, L.}
\newblock \bibinfo{journal}{\bibinfo{title}{Social {{Confinement}} and
  {{Mesoscopic Localization}} of {{Epidemics}} on {{Networks}}}}.
\newblock {\emph{\JournalTitle{Phys. Rev. Lett.}}}
  \textbf{\bibinfo{volume}{126}}, \bibinfo{pages}{098301},
  \doiprefix\url{10.1103/PhysRevLett.126.098301} (\bibinfo{year}{2021}).

\bibitem{pastor2001epidemic}
\bibinfo{author}{Pastor-Satorras, R.} \& \bibinfo{author}{Vespignani, A.}
\newblock \bibinfo{journal}{\bibinfo{title}{Epidemic spreading in scale-free
  networks}}.
\newblock {\emph{\JournalTitle{Phys. Rev. Lett.}}}
  \textbf{\bibinfo{volume}{86}}, \bibinfo{pages}{3200} (\bibinfo{year}{2001}).

\bibitem{stonge2018phase}
\bibinfo{author}{St-Onge, G.}, \bibinfo{author}{Young, J.-G.},
  \bibinfo{author}{Laurence, E.}, \bibinfo{author}{Murphy, C.} \&
  \bibinfo{author}{Dub{\'e}, L.~J.}
\newblock \bibinfo{journal}{\bibinfo{title}{Phase transition of the
  susceptible-infected-susceptible dynamics on time-varying configuration model
  networks}}.
\newblock {\emph{\JournalTitle{Phys. Rev. E}}} \textbf{\bibinfo{volume}{97}},
  \bibinfo{pages}{022305} (\bibinfo{year}{2018}).

\bibitem{dorogovtsev2003spectra}
\bibinfo{author}{Dorogovtsev, S.~N.}, \bibinfo{author}{Goltsev, A.~V.},
  \bibinfo{author}{Mendes, J. F.~F.} \& \bibinfo{author}{Samukhin, A.~N.}
\newblock \bibinfo{journal}{\bibinfo{title}{Spectra of complex networks}}.
\newblock {\emph{\JournalTitle{Phys. Rev. E}}} \textbf{\bibinfo{volume}{68}},
  \bibinfo{pages}{046109}, \doiprefix\url{10.1103/PhysRevE.68.046109}
  (\bibinfo{year}{2003}).

\bibitem{goltsev2012localization}
\bibinfo{author}{Goltsev, A.~V.}, \bibinfo{author}{Dorogovtsev, S.~N.},
  \bibinfo{author}{Oliveira, J.~G.} \& \bibinfo{author}{Mendes, J. F.~F.}
\newblock \bibinfo{journal}{\bibinfo{title}{Localization and {{Spreading}} of
  {{Diseases}} in {{Complex Networks}}}}.
\newblock {\emph{\JournalTitle{Phys. Rev. Lett.}}}
  \textbf{\bibinfo{volume}{109}}, \bibinfo{pages}{128702},
  \doiprefix\url{10.1103/PhysRevLett.109.128702} (\bibinfo{year}{2012}).

\bibitem{pastor-satorras2018eigenvector}
\bibinfo{author}{{Pastor-Satorras}, R.} \& \bibinfo{author}{Castellano, C.}
\newblock \bibinfo{journal}{\bibinfo{title}{Eigenvector {{Localization}} in
  {{Real Networks}} and {{Its Implications}} for {{Epidemic Spreading}}}}.
\newblock {\emph{\JournalTitle{J. Stat. Phys.}}}
  \textbf{\bibinfo{volume}{173}}, \bibinfo{pages}{1110--1123},
  \doiprefix\url{10.1007/s10955-018-1970-8} (\bibinfo{year}{2018}).

\bibitem{st-onge2021master}
\bibinfo{author}{St-Onge, G.}, \bibinfo{author}{Thibeault, V.},
  \bibinfo{author}{Allard, A.}, \bibinfo{author}{Dub{\'e}, L.~J.} \&
  \bibinfo{author}{H{\'e}bert-Dufresne, L.}
\newblock \bibinfo{journal}{\bibinfo{title}{Master equation analysis of
  mesoscopic localization in contagion dynamics on higher-order networks}}.
\newblock {\emph{\JournalTitle{Phys. Rev. E}}} \textbf{\bibinfo{volume}{103}},
  \bibinfo{pages}{032301} (\bibinfo{year}{2021}).

\bibitem{hebert-dufresne2016multiscale}
\bibinfo{author}{{H{\'e}bert-Dufresne}, L.}, \bibinfo{author}{Grochow, J.~A.}
  \& \bibinfo{author}{Allard, A.}
\newblock \bibinfo{journal}{\bibinfo{title}{Multi-scale structure and
  topological anomaly detection via a new network statistic: {{The}} onion
  decomposition}}.
\newblock {\emph{\JournalTitle{Sci. Rep.}}} \textbf{\bibinfo{volume}{6}},
  \bibinfo{pages}{31708}, \doiprefix\url{10.1038/srep31708}
  (\bibinfo{year}{2016}).

\bibitem{hebert-dufresne2010propagation}
\bibinfo{author}{{H{\'e}bert-Dufresne}, L.}, \bibinfo{author}{No{\"e}l, P.-A.},
  \bibinfo{author}{Marceau, V.}, \bibinfo{author}{Allard, A.} \&
  \bibinfo{author}{Dub{\'e}, L.~J.}
\newblock \bibinfo{journal}{\bibinfo{title}{Propagation dynamics on networks
  featuring complex topologies}}.
\newblock {\emph{\JournalTitle{Phys. Rev. E}}} \textbf{\bibinfo{volume}{82}},
  \bibinfo{pages}{036115}, \doiprefix\url{10.1103/PhysRevE.82.036115}
  (\bibinfo{year}{2010}).

\bibitem{allard2019percolation}
\bibinfo{author}{Allard, A.} \& \bibinfo{author}{{H{\'e}bert-Dufresne}, L.}
\newblock \bibinfo{journal}{\bibinfo{title}{Percolation and the {{Effective
  Structure}} of {{Complex Networks}}}}.
\newblock {\emph{\JournalTitle{Phys. Rev. X}}} \textbf{\bibinfo{volume}{9}},
  \bibinfo{pages}{011023}, \doiprefix\url{10.1103/PhysRevX.9.011023}
  (\bibinfo{year}{2019}).

\bibitem{redner2019realityinspired}
\bibinfo{author}{Redner, S.}
\newblock \bibinfo{journal}{\bibinfo{title}{Reality-inspired voter models:
  {{A}} mini-review}}.
\newblock {\emph{\JournalTitle{C. R. Phys.}}} \textbf{\bibinfo{volume}{20}},
  \bibinfo{pages}{275--292}, \doiprefix\url{10.1016/j.crhy.2019.05.004}
  (\bibinfo{year}{2019}).

\bibitem{hebert-dufresne2019smeared}
\bibinfo{author}{{H{\'e}bert-Dufresne}, L.} \& \bibinfo{author}{Allard, A.}
\newblock \bibinfo{journal}{\bibinfo{title}{Smeared phase transitions in
  percolation on real complex networks}}.
\newblock {\emph{\JournalTitle{Phys. Rev. Research}}}
  \textbf{\bibinfo{volume}{1}}, \bibinfo{pages}{013009},
  \doiprefix\url{10.1103/PhysRevResearch.1.013009} (\bibinfo{year}{2019}).

\bibitem{pennings1973measures}
\bibinfo{author}{Pennings, J.}
\newblock \bibinfo{journal}{\bibinfo{title}{Measures of organizational
  structure: A methodological note}}.
\newblock {\emph{\JournalTitle{American Journal of Sociology}}}
  \textbf{\bibinfo{volume}{79}}, \bibinfo{pages}{686--704}
  (\bibinfo{year}{1973}).

\bibitem{Iacopini2019simplicial}
\bibinfo{author}{Iacopini, I.}, \bibinfo{author}{Petri, G.},
  \bibinfo{author}{Barrat, A.} \& \bibinfo{author}{Latora, V.}
\newblock \bibinfo{journal}{\bibinfo{title}{Simplicial models of social
  contagion}}.
\newblock {\emph{\JournalTitle{Nature Communications}}}
  \textbf{\bibinfo{volume}{10}}, \bibinfo{pages}{1--9} (\bibinfo{year}{2019}).

\bibitem{st-onge2021influential}
\bibinfo{author}{{St-Onge}, G.}, \bibinfo{author}{Iacopini, I.} \&
  \bibinfo{author}{Latora, V. e.~a.}
\newblock \bibinfo{journal}{\bibinfo{title}{Influential groups for seeding and
  sustaining nonlinear contagion in heterogeneous hypergraphs}}.
\newblock {\emph{\JournalTitle{Commun Phys.}}} \textbf{\bibinfo{volume}{5}},
  \doiprefix\url{10.1038/s42005-021-00788-w} (\bibinfo{year}{2022}).

\bibitem{szulanski2016stickiness}
\bibinfo{author}{Szulanski, G.}, \bibinfo{author}{Ringov, D.} \&
  \bibinfo{author}{Jensen, R.~J.}
\newblock \bibinfo{journal}{\bibinfo{title}{Overcoming {{Stickiness}}: {{How}}
  the {{Timing}} of {{Knowledge Transfer Methods Affects Transfer
  Difficulty}}}}.
\newblock {\emph{\JournalTitle{Organization Science}}}
  \textbf{\bibinfo{volume}{27}}, \bibinfo{pages}{304--322},
  \doiprefix\url{10.1287/orsc.2016.1049} (\bibinfo{year}{2016}).

\bibitem{edmondson2018crossboundary}
\bibinfo{author}{Edmondson, A.~C.} \& \bibinfo{author}{Harvey, J.-F.}
\newblock \bibinfo{journal}{\bibinfo{title}{Cross-boundary teaming for
  innovation: {{Integrating}} research on teams and knowledge in
  organizations}}.
\newblock {\emph{\JournalTitle{Human Resource Management Review}}}
  \textbf{\bibinfo{volume}{28}}, \bibinfo{pages}{347--360},
  \doiprefix\url{10.1016/j.hrmr.2017.03.002} (\bibinfo{year}{2018}).

\bibitem{kensbock2021mental}
\bibinfo{author}{Kensbock, J.~M.}, \bibinfo{author}{Alk{\ae}rsig, L.} \&
  \bibinfo{author}{Lomberg, C.}
\newblock \bibinfo{journal}{\bibinfo{title}{The {{Epidemic}} of {{Mental
  Disorders}} in {{Business}}\textemdash{{How Depression}}, {{Anxiety}}, and
  {{Stress Spread}} across {{Organizations}} through {{Employee Mobility}}}}.
\newblock {\emph{\JournalTitle{Administrative Science Quarterly}}}
  \bibinfo{pages}{00018392211014819}, \doiprefix\url{10.1177/00018392211014819}
  (\bibinfo{year}{2021}).

\bibitem{noschese2013}
\bibinfo{author}{Noschese, S.}, \bibinfo{author}{Pasquini, L.} \&
  \bibinfo{author}{Reichel, L.}
\newblock \bibinfo{journal}{\bibinfo{title}{Tridiagonal toeplitz matrices:
  properties and novel applications}}.
\newblock {\emph{\JournalTitle{Numer. Linear Algebra Appl.}}}
  \textbf{\bibinfo{volume}{20}}, \bibinfo{pages}{302--326},
  \doiprefix\url{https://doi.org/10.1002/nla.1811} (\bibinfo{year}{2013}).
\newblock \eprint{https://onlinelibrary.wiley.com/doi/pdf/10.1002/nla.1811}.

\end{thebibliography}
\end{document}